\documentclass[conference]{IEEEtran}
\IEEEoverridecommandlockouts
\usepackage{cite}
\usepackage{amsmath,amssymb,amsfonts}
\usepackage{graphicx}
\usepackage{url}
\usepackage{booktabs}
\usepackage{xcolor}
\usepackage{algorithm}
\usepackage{algpseudocode}
\definecolor{forestgreen}{RGB}{34,139,34}

\usepackage{tikz}
\usetikzlibrary{positioning,shapes.geometric,calc}
\usepackage{placeins}
\newtheorem{definition}{Definition}
\newtheorem{theorem}{Theorem}
\newtheorem{proposition}{Proposition}
\newtheorem{exampleinner}{Example}
\newenvironment{example}{\begin{exampleinner}\upshape}{\hfill$\triangleleft$\end{exampleinner}}
\newcommand{\F}{\mathbb{F}_2}
\newcommand{\DMAP}{\mathrm{DMAP}}
\newcommand{\MAP}{\mathrm{MAP}}

\newcommand{\sym}[1]{\{#1\}}
\DeclareMathOperator{\spn}{span}
\newcommand{\pl}[1]{\mathsf{#1}}
\DeclareMathOperator*{\argmax}{arg\,max}
\DeclareMathOperator*{\maxs}{max^*}

\tikzset{
  chk/.style={draw,rectangle,minimum size=3.6mm,inner sep=0pt,fill=gray!15,font=\scriptsize},
  obs/.style={draw,diamond,minimum size=4.6mm,inner sep=0pt,fill=blue!8,font=\scriptsize},
  bv/.style={draw,circle,minimum size=2.6mm,inner sep=0pt},
  bvm/.style={draw,circle,minimum size=2.6mm,inner sep=0pt,fill=gray!45},
  v2/.style={draw,circle,minimum size=5.2mm,inner sep=0pt,fill=cyan!15,font=\scriptsize},
  v4/.style={draw,circle,minimum size=6.4mm,inner sep=0pt,fill=orange!25,font=\scriptsize},
  v1/.style={draw,circle,minimum size=4.2mm,inner sep=0pt,fill=white,font=\scriptsize},
  lab/.style={font=\tiny,fill=white,inner sep=0.6pt},
}

\begin{document}

\title{Non-Binary Detector Error Models \vspace{-2mm}}

\author{
	\IEEEauthorblockN{Stefano Tinelli$^{\dagger\ddagger}$, Francisco L\'azaro$^\dagger$, Francisco Garc\'ia Herrero$^\ddagger$}
	\IEEEauthorblockA{$^\dagger$Institute of Communications and Navigation, German Aerospace Center (\{Stefano.Tinelli,Francisco.LazaroBlasco\}@dlr.de) \\
		$^\ddagger$Universidad Complutense de Madrid, Spain (francg18@ucm.es)}
	\vspace{-0.4cm}
	\thanks{The work of Stefano Tinelli and Francisco L\'azaro is part of the HESC lighthouse project within Munich Quantum Valley initiative and is supported by the Bavarian state government with funds from the Hightech Agenda Bavaria.
	}
}
\date{}
\maketitle
\thispagestyle{empty}
\pagestyle{empty}

\begin{abstract}
A key challenge in fault-tolerant quantum error correction (QEC) is that the syndrome extraction circuits are themselves faulty. Such circuit-level noise is commonly described by a detector error model (DEM). In Stim and in almost all related works, DEMs are binary: every non-binary circuit error (e.g., CNOT gate), is decomposed into independent binary error mechanisms. 
In this work, we introduce non-binary DEMs, in which each fault location of the syndrome extraction circuit is modeled by one random variable over its original alphabet, e.g., the 16 two-qubit Pauli errors after a CNOT gate. We also propose a split-merge-regroup transformation that builds a compact non-binary DEM from any Clifford circuit by merging the error components that flip the same detectors and observables.
Non-binary DEMs represent any Pauli channel exactly, whereas binary DEMs are exact only for channels that factorize into independent binary parts. When the channels factorize, we prove that binary and non-binary DEMs are equivalent under degenerate quantum maximum-a-posteriori (MAP) decoding, but not under quantum MAP decoding. Although the variables are non-binary, belief propagation (BP) still exchanges a single scalar message per edge. 
Our numerical results with surface, color and bivariate bicycle codes show that non-binary DEMs have two main advantages. First, their Tanner graphs have four to five times fewer edges, which is advantageous for hardware implementations. Second, their performance under BP, with and without OSD post-processing, is better than that of their binary counterparts, and BP with OSD runs up to $9.4\times$ faster in a CPU. 
\end{abstract}

\section{Introduction}

In real quantum hardware, errors affect not only the data qubits but also the ancilla qubits, the gates, the resets and the measurements of the syndrome extraction circuit~\cite{dennis2002topological,fowler2012surface}.
The standard abstraction for decoding under such \emph{circuit-level} noise is the \emph{detector error model} (DEM), introduced with the simulator Stim~\cite{gidney2021stim} and studied systematically in~\cite{derks2025designing}.
A DEM is a list of independent binary fault mechanisms, together with their probabilities and the detectors and logical observables that each of them flips. It is thus a sparse parity-check matrix with priors, i.e., a Tanner graph, and classical decoders such as minimum-weight perfect matching (MWPM)~\cite{higgott2022pymatching} or belief propagation (BP)~\cite{mackay2004sparse,poulin2008iterative} and its refinements~\cite{panteleev2021degenerate,roffe2020decoding,muller2025relay} can be directly applied to it.

However, the errors affecting quantum circuits are not binary. For example, a Pauli error on an idling qubit has four possible outcomes ($I,X,Y,Z$), and a faulty CNOT gate has sixteen.
To obtain a binary DEM, every such error channel is first written as the XOR of \emph{independent} binary mechanisms, three for a single-qubit ($X,Y,Z$) and fifteen for a two-qubit depolarizing channel.
This splitting creates many 4-cycles in the Tanner graph (see Proposition~\ref{prop:size}).
Subsequently, mechanisms that flip the same detectors and observables, possibly from different locations, are merged, which removes some of these short cycles, but not those inside a fault location.
A more natural alternative is to employ one, possibly non-binary, random variable per fault location, e.g., the 16 two-qubit Pauli errors after a CNOT gate. 
In the context of classical channel codes, replacing binary by non-binary variables, as in low-density parity-check codes over $\mathrm{GF}(2^r)$, turns the scalar BP messages into vectors of length $2^r-1$ \cite{davey1998low}.
However, this is not the case for quantum codes since the detectors (check-nodes) stay binary (see \cite{kuo2020refined,lai2021log,ostrev2024classical}). 
This idea was extended to circuit-level noise in~\cite{kuo2026streaming}, which models circuit-level noise with non-binary random variables and introduces \emph{probabilistic error consolidation} to mitigate degeneracy and short cycles in a streaming BP decoder. In contrast, we study the DEM itself.

Our contributions are the following.

\begin{itemize}
\item We define \emph{non-binary DEMs}, in which every fault location is associated with one random variable over its native alphabet $\F^{r}$, $r\in\{1,2,4\}$ (i.e., binary, quaternary or 16-ary), with its exact prior. They generalize the DEMs of Stim, which are the special case of binary variables.
\item We introduce a transformation that splits all locations into binary parts, merges parts that flip the same detectors and observables, and regroups the merged parts into non-binary variables (of rank $r\le4$). We prove that this construction is \emph{exact} whenever the channels factorize into independent binary parts, as depolarizing channels do, so that binary and non-binary DEMs are equivalent under degenerate quantum MAP decoding. Moreover, non-binary DEMs represent any Pauli channel exactly, whereas binary DEMs can in general only approximate channels that do not factorize. Under quantum MAP decoding, the two DEMs differ.
\item We compare BP, with and without ordered statistics decoding (OSD), on binary and non-binary DEMs on identical shots, and show that the non-binary DEM yields a lower logical error rate for surface, color and bivariate bicycle (BB) code instances. 
Additionally, our results show that non-binary DEMs have $5.9$--$8\times$ fewer variable nodes and $4.3$--$5.3\times$ fewer edges than the binary one for the codes we consider, and make BP with OSD up to $9.4\times$ faster in a CPU.
\end{itemize}

\vspace{-2mm}
\section{Preliminaries}\label{sec:prelim}
\vspace{-1mm}
\subsection{Detectors, observables and fault locations}

We consider Clifford circuits with Pauli noise. A \emph{detector} is a parity of measurement outcomes that is deterministic in the absence of noise, and a (logical) \emph{observable} is a parity (linear combination) of measurement outcomes whose value we want to protect~\cite{gidney2021stim,derks2025designing}. 
We denote the number of detectors by $m$ and the number of observables by $k$. A Pauli error that occurs somewhere in the circuit flips a subset of the detectors and of the observables. We call this subset the \emph{symptom} of the error and write it either as a set, e.g., $\sym{D_4,L_1}$, or as a vector $(\mathbf{s},\boldsymbol{\ell})\in\F^{m}\times\F^{k}$. The symptom of a product of Pauli errors is the sum (over $\F$) of their symptoms. In a shot of the experiment, the errors that occur produce one total symptom. Its detector part $\mathbf{s}$ is the \emph{syndrome}, which the decoder observes, and its observable part $\boldsymbol{\ell}$ is the \emph{logical flip}, which the decoder has to estimate. A \emph{fault location} is a noisy operation: a single- or two-qubit channel, or a bit flip after a reset or before a measurement.

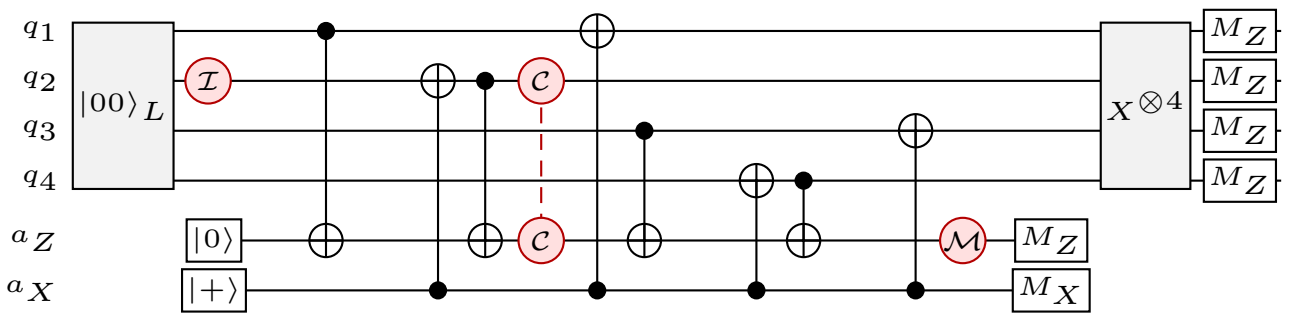
\begin{figure}[t]
\centering
\begin{tikzpicture}[x=0.78cm,y=0.33cm,font=\scriptsize,
  ctl/.style={circle,fill,inner sep=0pt,minimum size=1.2mm},
  tgt/.style={draw,circle,inner sep=0pt,minimum size=2.2mm,path picture={\draw (path picture bounding box.north)--(path picture bounding box.south) (path picture bounding box.west)--(path picture bounding box.east);}},
  gate/.style={draw,fill=white,inner sep=0.8pt,minimum height=2.8mm,font=\tiny},
  flt/.style={draw=red!70!black,fill=red!12,circle,inner sep=0pt,minimum size=3.0mm,font=\tiny}]
\foreach \y/\n in {0/q_1,-1/q_2,-2/q_3,-3/q_4} { \node[anchor=east,font=\tiny] at (0,\y) {$\n$}; \draw (0,\y)--(10.2,\y); }
\node[anchor=east,font=\tiny] at (0,-4.2) {$a_Z$}; \node[anchor=east,font=\tiny] at (0,-5.2) {$a_X$};
\draw (1.15,-4.2)--(8.25,-4.2); \draw (1.15,-5.2)--(8.25,-5.2);
\node[gate,minimum height=1.1cm,fill=gray!10] at (0.38,-1.5) {$|00\rangle_L$};
\node[gate] at (1.15,-4.2) {$|0\rangle$}; \node[gate] at (1.15,-5.2) {$|{+}\rangle$};
\node[flt] at (1.1,-1) {$\mathcal{I}$};
\foreach \t/\zq/\xq in {1/0/-1,2/-1/0,3/-2/-3,4/-3/-2} {
  \pgfmathsetmacro{\xz}{0.75+1.35*\t} \pgfmathsetmacro{\xx}{\xz+0.95}
  \draw (\xz,\zq)--(\xz,-4.2); \node[ctl] at (\xz,\zq) {}; \node[tgt] at (\xz,-4.2) {};
  \draw (\xx,-5.2)--(\xx,\xq); \node[ctl] at (\xx,-5.2) {}; \node[tgt] at (\xx,\xq) {};
}
\draw[red!70!black,densely dashed] (3.925,-1)--(3.925,-4.2);
\node[flt] at (3.925,-1) {$\mathcal{C}$}; \node[flt] at (3.925,-4.2) {$\mathcal{C}$};
\node[flt] at (7.5,-4.2) {$\mathcal{M}$};
\node[gate] at (8.25,-4.2) {$M_Z$}; \node[gate] at (8.25,-5.2) {$M_X$};
\node[gate,minimum height=1.1cm,fill=gray!10] at (9.05,-1.5) {$X^{\otimes4}$};
\foreach \y in {0,-1,-2,-3} { \node[gate] at (9.85,\y) {$M_Z$}; }
\end{tikzpicture}
\caption{Circuit of Example~\ref{ex:circuit} with its three fault locations (red). Preparation and the last $X^{\otimes4}$ measurement are noiseless. $D_1$: outcome of $a_Z$; $D_4$: $a_Z$ $\oplus$ readout of $q_1,\dots,q_4$; $D_2$ ($D_3$): $a_X$ $\oplus$ first (last) $X^{\otimes4}$.}
\label{fig:circuit} \vspace{-3mm}
\end{figure}

\begin{example}\label{ex:circuit} 
Consider one round of syndrome measurement of the $[[4,2,2]]$ code (Fig.~\ref{fig:circuit}) with stabilizers $Z^{\otimes4}$ and $X^{\otimes4}$, measured with two ancillas,  $a_Z$ and $a_X$, followed by the readout of the data qubits $q_1,\dots,q_4$. 
The circuit has two $Z$-type detectors $D_1,D_4$ (they detect $X$ errors), two $X$-type detectors $D_2,D_3$, and an observable $L_1$ (the readout parity of $q_1$ and $q_2$). 
Although the circuit has many more fault locations, for illustration we concentrate only on three of them: a single-qubit depolarizing channel $\mathcal{I}$ on $q_2$ before the round, a two-qubit depolarizing channel $\mathcal{C}$ after the CNOT from $q_2$ to $a_Z$, and a measurement flip $\mathcal{M}$ of $a_Z$. 
Table~\ref{tab:gens} lists the symptoms of their $X$ and $Z$ errors ($Y$ errors follow by addition). For instance, an $X$ error on $q_2$ right after the CNOT (fault location $\mathcal{C}$) flips $D_4$ and $L_1$, while a $Z$ error on $a_Z$ at the same location flips $D_2$ and $D_3$.
\end{example}

\vspace{-2mm}
\begin{table}[t]
\centering
\caption{Symptoms of the $X$ and $Z$ errors at the three fault locations of Example~\ref{ex:circuit}.%
}
\label{tab:gens}
\footnotesize
\setlength{\tabcolsep}{3.6pt}
\begin{tabular}{@{}c|cc|cccc|c@{}}
\toprule
 & \multicolumn{2}{c|}{$\mathcal{I}$ (on $q_2$)} & \multicolumn{4}{c|}{$\mathcal{C}$ (after CNOT $q_2\to a_Z$)} & $\mathcal{M}$\\
 & $X_{q_2}$ & $Z_{q_2}$ & $X_{q_2}$ & $Z_{q_2}$ & $X_{a_Z}$ & $Z_{a_Z}$ & flip\\
 & $x$ & $z$ & $x_c$ & $z_c$ & $x_t$ & $z_t$ & $b$\\
\midrule
$D_1$ & 1&0 & 0&0&1&0 & 1\\
$D_2$ & 0&1 & 0&0&0&1 & 0\\
$D_3$ & 0&0 & 0&1&0&1 & 0\\
$D_4$ & 0&0 & 1&0&1&0 & 1\\
\midrule
$L_1$ & 1&0 & 1&0&0&0 & 0\\
\bottomrule
\end{tabular}
\vspace{-4mm}
\end{table}

\subsection{Binary detector error models}
\begin{definition}[Binary DEM]
A binary DEM is a list of $n$ mechanisms $(p_j,\mathbf{h}_j,\mathbf{l}_j)$, $j=1,\dots,n$, with $p_j\in(0,\frac12)$, $\mathbf{h}_j\in\F^{m}$ and $\mathbf{l}_j\in\F^{k}$. The mechanisms $e_j\sim\mathrm{Bern}(p_j)$ are independent, and $\mathbf{s}=H\mathbf{e}$, $\boldsymbol{\ell}=L\mathbf{e}$, with $H=[\mathbf{h}_1\cdots\mathbf{h}_n]$ and $L=[\mathbf{l}_1\cdots\mathbf{l}_n]$.
\end{definition}

A fault location is turned into binary mechanisms by writing its channel as the XOR of \emph{independent} binary parts. 
For example, a single-qubit depolarizing channel of strength $p$ is the XOR of three independent parts ($X$, $Y$ and $Z$), each with probability $q$, where $(1-2q)^2=1-\frac43p$. 
Similarly, a two-qubit depolarizing channel, which models an erroneous CNOT gate, is the XOR of fifteen independent  parts. We say that such channels \emph{factorize}. Depolarizing channels factorize, but not every Pauli channel does (see the remark after Theorem~\ref{thm:dmap}). 

After splitting each fault location into its binary parts, it is convenient to merge equivalent errors (duplicated columns), as it is done in Stim~\cite{gidney2021stim}. This is akin to representing two independent Bernoulli variables with parameters $q_1,q_2$ that flip the same detectors by their XOR, i.e., a single Bernoulli random variable with parameter $q_1+q_2-2q_1q_2$. 
We denote the joint distribution of $(\mathbf{s},\boldsymbol{\ell})$ by $P(\mathbf{s},\boldsymbol{\ell})$.

\begin{example}\label{ex:binary} 
Splitting the three locations of Example~\ref{ex:circuit} yields $3+15+1=19$ parts. Four pairs of them have equal symptoms: $\mathcal{M}$ has the symptom of the part $X_{a_Z}$ of $\mathcal{C}$, and the parts $X$, $Z$ and $Y$ of $\mathcal{I}$ have the symptoms of the parts $X_{q_2}X_{a_Z}$, $Z_{q_2}Z_{a_Z}$ and $Y_{q_2}Y_{a_Z}$ of $\mathcal{C}$. The binary DEM therefore has $n=15$ mechanisms. Its Tanner graph, shown in Fig.~\ref{fig:tanner}(a), has 32 edges and 36 4-cycles.
\end{example}

\subsection{Degenerate quantum MAP and quantum MAP decoding}

Let us remark that faults with the same syndrome are indistinguishable, whereas faults that differ by an error with an empty symptom are equivalent. 
Hence, the optimal decoder estimates the logical flip, not the fault \cite{iyer2015hardness,pryadko2020ml}. Formally, given $\mathbf{s}$, the \emph{degenerate quantum MAP} (DMAP) decoder outputs
\vspace{-1mm}
\begin{equation}
\hat{\boldsymbol{\ell}}_{\DMAP}(\mathbf{s})=\argmax_{\boldsymbol{\ell}}P(\mathbf{s},\boldsymbol{\ell}).
\end{equation}
In contrast, the \emph{quantum MAP} decoder outputs the most likely error. Consider a model with fault variables $\mathbf{x}$, prior $P(\mathbf{x})$ and symptom map $\mathbf{x}\mapsto(\mathbf{s}(\mathbf{x}),\boldsymbol{\ell}(\mathbf{x}))$. Given $\mathbf{s}$, the quantum MAP decoder outputs $\hat{\boldsymbol{\ell}}_{\MAP}(\mathbf{s})=\boldsymbol{\ell}(\hat{\mathbf{x}})$, where
\vspace{-1mm}
\begin{equation}
	\hat{\mathbf{x}}=\argmax_{\mathbf{x}:\,\mathbf{s}(\mathbf{x})=\mathbf{s}}P(\mathbf{x}).
\end{equation}

\vspace{-2mm}
\section{Non-binary detector error models}\label{sec:nbdem}
\subsection{Definition}
\begin{definition}[Non-binary DEM]
A non-binary DEM is a list of $N$ columns $(\pi_c,H_c,L_c)$, $c=1,\dots,N$. Column $c$ is a random variable $u_c\in\F^{r_c}$ with prior distribution $\pi_c$, and $H_c\in\F^{m\times r_c}$, $L_c\in\F^{k\times r_c}$ are matrices. The variables $u_1,\dots,u_N$ are independent, and
\vspace{-1mm}
\begin{equation}\label{eq:nbdem}
\mathbf{s}=\sum_{c=1}^{N}H_cu_c,\qquad \boldsymbol{\ell}=\sum_{c=1}^{N}L_cu_c.\vspace{-1mm}
\end{equation}
\end{definition}

The Tanner graph associated to a non-binary DEM has one variable node per column and one (binary) check node per detector, and its edges are labelled.
In particular, there is an edge between detector $d$ and column $c$ if row $d$ of $H_c$, denoted by $\mathbf{h}_{d,c}\in\F^{r_c}$, is nonzero, and this row determines the \emph{label}.
Following \cite{kuo2020refined,kuo2026streaming}, we label the edges connected to quaternary and 16-ary variable nodes using Pauli operators.
A quaternary variable $u_c=(x,z)$ connected to a check node through an edge with $\mathbf{h}_{d,c}=(h_1,h_2)$ will flip the detector if and only if $\langle\mathbf{h}_{d,c},u_c\rangle=h_1x\oplus h_2z=1$,
	i.e., if and only if the error $X^{x}Z^{z}$ anticommutes with $X^{h_2}Z^{h_1}$.
Thus, we set\footnote{The components are swapped since an $X$ error anticommutes with $Z$, and a $Z$ error with $X$.} $P_{d,c}=X^{h_2}Z^{h_1}$.
Similarly, for a 16-ary node, $u_c=(x_1,z_1,x_2,z_2)$ and $\mathbf{h}_{d,c}=(h_1,h_2,h_3,h_4)$, we set $P_{d,c}=X^{h_2}Z^{h_1}\otimes X^{h_4}Z^{h_3}$. In both cases, the detector is flipped if and only if the error anticommutes with $P_{d,c}$, and edges to binary nodes carry no label.
For a quaternary node, the labels $\pl{Z}$, $\pl{X}$ and $\pl{Y}$ (binary $(1,0)$, $(0,1)$ and $(1,1)$) mean that the detector is flipped by $X$ and $Y$, by $Z$ and $Y$, or by $X$ and $Z$, respectively. For a 16-ary node, e.g., a detector with label $\pl{IZ}$ on a CNOT location is flipped by an $X$ or $Y$ error on the second qubit, and one with label $\pl{XX}$ is flipped if exactly one of the two qubits has a $Z$ or $Y$ error. 
Hence, the labelled Tanner graph generalizes the quaternary parity-check matrix of a stabilizer code~\cite{babar2015fifteen,kuo2020refined,ostrev2024classical}, written over the fault locations of a circuit instead of the data qubits.

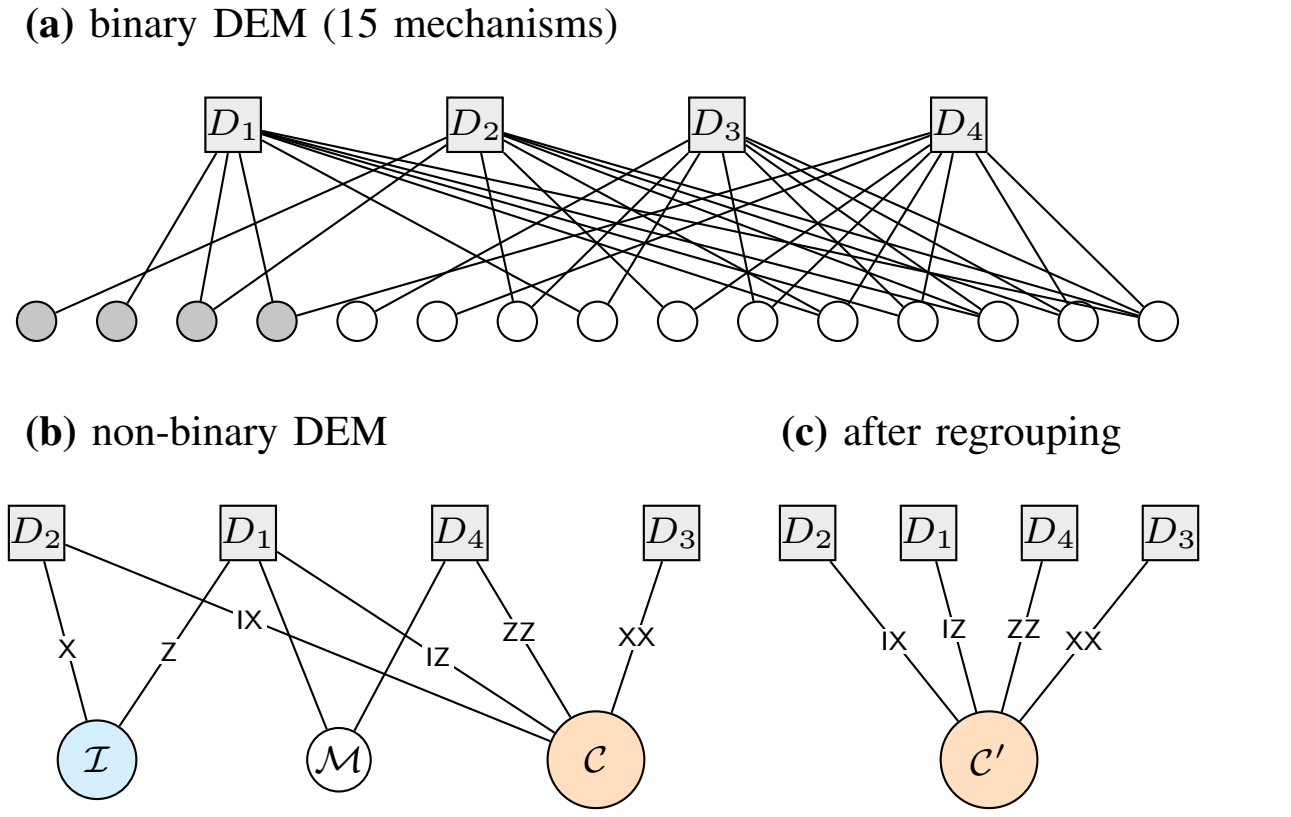
\begin{figure}[t]
\centering
\begin{tikzpicture}[font=\scriptsize]
\begin{scope}
\node[font=\footnotesize,anchor=west] at (-0.2,2.25) {\textbf{(a)} binary DEM (15 mechanisms)};
\node[chk] (b0) at (1.30,1.6) {$D_1$}; \node[chk] (b1) at (2.90,1.6) {$D_2$}; \node[chk] (b2) at (4.50,1.6) {$D_3$}; \node[chk] (b3) at (6.10,1.6) {$D_4$};
\foreach \i/\t/\nb in {0/bvm/{1},1/bvm/{0},2/bvm/{0,1},3/bvm/{0,3},4/bv/{2},5/bv/{3},6/bv/{1,2},7/bv/{0,2},8/bv/{1,3},9/bv/{2,3},10/bv/{0,1,3},11/bv/{0,2,3},12/bv/{0,1,2},13/bv/{1,2,3},14/bv/{0,1,2,3}} {
  \node[\t] (v\i) at (0.53*\i,0.3) {};
  \foreach \j in \nb { \draw (v\i)--(b\j); }
}
\end{scope}
\begin{scope}[yshift=-2.7cm]
\node[font=\footnotesize,anchor=west] at (-0.2,2.25) {\textbf{(b)} non-binary DEM};
\node[chk] (d2) at (0,1.6) {$D_2$}; \node[chk] (d1) at (1.4,1.6) {$D_1$}; \node[chk] (d4) at (2.8,1.6) {$D_4$}; \node[chk] (d3) at (4.2,1.6) {$D_3$};
\node[v2] (I) at (0.4,0.1) {$\mathcal{I}$}; \node[v1] (M) at (2.0,0.1) {$\mathcal{M}$}; \node[v4] (C) at (3.7,0.1) {$\mathcal{C}$};
\draw (I)--node[lab,pos=0.45]{$\pl{X}$}(d2); \draw (I)--node[lab,pos=0.45]{$\pl{Z}$}(d1);
\draw (M)--(d1); \draw (M)--(d4);
\draw (C)--node[lab,pos=0.62]{$\pl{IX}$}(d2); \draw (C)--node[lab,pos=0.42]{$\pl{IZ}$}(d1); \draw (C)--node[lab,pos=0.55]{$\pl{ZZ}$}(d4); \draw (C)--node[lab,pos=0.5]{$\pl{XX}$}(d3);
\end{scope}
\begin{scope}[xshift=5.1cm,yshift=-2.7cm]
\node[font=\footnotesize,anchor=west] at (-0.3,2.25) {\textbf{(c)} after regrouping};
\node[chk] (e2) at (0,1.6) {$D_2$}; \node[chk] (e1) at (0.8,1.6) {$D_1$}; \node[chk] (e4) at (1.6,1.6) {$D_4$}; \node[chk] (e3) at (2.4,1.6) {$D_3$};
\node[v4] (C') at (1.2,0.1) {$\mathcal{C}'$};
\draw (C')--node[lab,pos=0.5]{$\pl{IX}$}(e2); \draw (C')--node[lab,pos=0.55]{$\pl{IZ}$}(e1); \draw (C')--node[lab,pos=0.55]{$\pl{ZZ}$}(e4); \draw (C')--node[lab,pos=0.5]{$\pl{XX}$}(e3);
\end{scope}
\end{tikzpicture}
\vspace{-3mm}
\caption{Tanner graphs of the three fault locations of Example~\ref{ex:circuit}. (a) Binary DEM; gray nodes are merged from parts of two locations. (b) Non-binary DEM with a quaternary ($\mathcal{I}$), a 16-ary ($\mathcal{C}$) and a binary ($\mathcal{M}$) node. The labels are Pauli operators: the detector is flipped if and only if the error anticommutes with the label. (c) After merging and regrouping (Example~\ref{ex:merge}). The observable is not shown.}
\label{fig:tanner}
\vspace{-6mm}
\end{figure}

\begin{example}[Non-binary DEM]\label{ex:nbdem}
The three locations of Example~\ref{ex:circuit} give a non-binary DEM with $N=3$ columns: a quaternary column $u_{\mathcal{I}}=(x,z)$, a 16-ary column $u_{\mathcal{C}}=(x_c,z_c,x_t,z_t)$, with the components on the control $q_2$ and on the target $a_Z$ of the CNOT, and a binary column $u_{\mathcal{M}}$. Their matrices $[H_c;L_c]$ are given in Table~\ref{tab:gens}, and Fig.~\ref{fig:tanner}(b) shows the Tanner graph. For instance, $D_3$ is connected to $\mathcal{C}$ with label $\pl{XX}$ (binary $0101$): it is flipped if $z_c\oplus z_t=1$. The graph has 3 variable nodes, 8 edges and two 4-cycles.
\end{example}

\subsection{Construction by splitting, merging and regrouping}\label{sec:merge}
So far we have described non-binary DEMs that introduce one (possibly non-binary) random variable to model noise at each location. 
However, this is a poor choice when using a suboptimal decoder such as belief propagation because parts of different locations frequently have equal symptoms. In our example, $\mathcal{M}$ and the three different Paulis of $\mathcal{I}$ are degenerate copies of parts of $\mathcal{C}$. Hence, they create 4-cycles in the Tanner graph, see Fig.~\ref{fig:tanner}(b). The following procedure can be used to transform a non-binary DEM into an equivalent one (see Theorem~\ref{thm:dmap}) while reducing the number of 4-cycles:
\begin{enumerate}
\item \emph{Split} every location $l$ into its independent binary parts $(q_i,v_i)$. A location whose channel does not factorize is not split and is kept as a column of its own, which is exact by Theorem~\ref{thm:dmap}(i).

\item \emph{Merge} parts with equal symptoms. Let $i$ and $h$ be two parts with equal symptom $v$ and probabilities $q_i$ and $q_h$. They are merged into a new part $j$ with symptom $v_j=v$ and probability $q_j=q_i+q_h-2q_iq_h$, the probability that exactly one of them occurs. This is repeated until all symptoms are distinct. For every merged part $j$, we keep track of its set of origin locations $O_j$, i.e., the locations whose parts were merged into $j$.

\item \emph{Assign} every merged part $j$ to the origin location of largest rank, $\mathrm{owner}(j)=\argmax_{l\in O_j}\rho_l$, where $\rho_l=\operatorname{rank}[H_l;L_l]$ is the rank of location $l$ in the non-binary DEM with one column per location, breaking ties by circuit order. 

\item \emph{Regroup.} Let $G_l$ be the set of parts assigned to $l$. Locations with $G_l=\emptyset$ are dropped. Column $l$ gets a basis $B_l=[H_l;L_l]$ of $\spn\{v_j:j\in G_l\}$, and $u_l$ the prior $\pi_l(u)=\Pr[\sum_{j\in G_l}e_ja_j=u]$, where $v_j=B_la_j$ and $e_j\sim\mathrm{Bern}(q_j)$ are independent. In practice, $\pi_l$ is a table of $2^{r_l}$ probabilities, initialized with all mass on $u=0$ and updated once per part $j\in G_l$ as $\pi_l(u)\gets(1-q_j)\,\pi_l(u)+q_j\,\pi_l(u\oplus a_j)$.
\end{enumerate}
At this point we would like to make two remarks. The first is that, after merging, one obtains Stim's binary DEM, which we verified numerically for the circuits of Section~\ref{sec:sims}. The second is that the rank-first rule lets large locations absorb degenerate parts of small ones, as illustrated in the following example.

\begin{example}[Merging]\label{ex:merge}
In Example~\ref{ex:nbdem}, $[H_{\mathcal{I}};L_{\mathcal{I}}]$ consists of the sums of columns 1,3 and 2,4 of $[H_{\mathcal{C}};L_{\mathcal{C}}]$, and $[H_{\mathcal{M}};L_{\mathcal{M}}]$ is column 3 (Table~\ref{tab:gens}). The 16-ary location $\mathcal{C}$ has the largest rank and claims all 15 merged parts. The result is a single 16-ary column $\mathcal{C}'$ with the labels of $\mathcal{C}$. Its parts are the 15 merged mechanisms of the binary DEM of Example~\ref{ex:binary}: the 15 parts of $\mathcal{C}$, four of which ($IX$, $XX$, $ZZ$, $YY$) absorbed $\mathcal{M}$ and the parts $X$, $Z$, $Y$ of $\mathcal{I}$. Its prior distribution $\pi_{\mathcal{C}'}$ is a table of 16 probabilities, one per two-qubit Pauli error. It starts with all mass on $u=0000$, i.e., on $II$, and is then updated once for each of the 15 parts $j$. With probability $q_j$ the part occurs and $u$ becomes $u\oplus a_j$, and with probability $1-q_j$ it does not occur and $u$ is unchanged.  Its Tanner graph, Fig.~\ref{fig:tanner}(c), has one node, four edges and no cycles.
\end{example}

Note that each merged part is claimed by exactly one column, and the detector support of a column is contained in the union of the supports of its parts. As a consequence, the Tanner graph of the non-binary DEM has at most as many variable nodes and edges as that of the binary DEM\footnote{In practice, the number of variable nodes and edges is actually considerably smaller, see Section~\ref{sec:sims}.}. 

\vspace{-1mm}
\subsection{Exactness and equivalence under DMAP}
\begin{theorem}[Exactness]\label{thm:dmap}
Consider a Clifford circuit with independent Pauli channels at its fault locations, and let $P(\mathbf{s},\boldsymbol{\ell})$ be the distribution of its detectors and observables.
(i) The non-binary DEM with one column per location, whose prior is the channel of that location, has distribution $P$.
(ii) If every channel factorizes, the binary DEM and any non-binary DEM obtained by grouping its merged parts into columns, in particular the output of the split-merge-regroup procedure of Section~\ref{sec:merge} for any assignment of the parts, also have distribution $P$.
In both cases, the DMAP decoders of all these models coincide and have the same LER.
\end{theorem}
\begin{IEEEproof}
(i) Clifford propagation is linear over $\F$, so an error $u_l$ at location $l$ has symptom $B_lu_l$, where the columns of $B_l=[H_l;L_l]$ are the symptoms of its $X$ and $Z$ components. The errors at different locations are independent, so $(\mathbf{s},\boldsymbol{\ell})=\sum_lB_lu_l$ with $u_l\sim\pi_l$, which is~\eqref{eq:nbdem}. (ii) Let $v_1,\dots,v_t\in\F^{m+k}$ be the symptoms of the independent parts and $e_1,\dots,e_t$ their Bernoulli variables. The total symptom is $(\mathbf{s},\boldsymbol{\ell})=\sum_ie_iv_i$. The binary DEM describes this random variable after merging parts with equal symptoms, which does not change its distribution. In the non-binary DEM, column $c$ describes $w_c=\sum_{i\in G_c}e_iv_i$, where $G_c$ is its set of parts. The $w_c$ depend on disjoint sets of independent variables and are independent, and $w_c=B_cu_c$, where the columns of $B_c=[H_c;L_c]$ are a basis of $\spn\{v_i:i\in G_c\}$ and $u_c$ has the prior distribution $\pi_c(u)=\Pr[\sum_{i\in G_c}e_ia_i=u]$, with $v_i=B_ca_i$. Both models thus describe $\sum_ie_iv_i$, and DMAP depends on $P(\mathbf{s},\boldsymbol{\ell})$ only.
\end{IEEEproof}
Note that the hypothesis of (ii) is necessary. For instance, the single-qubit channel with $\pi=(1-2a,a,a,0)$ for $I,X,Z,Y$ is not the XOR of independent parts $X',Z',Y'$, since the parity equations give $(1-2q_{Y'})^2=(1-2a)^2/(1-4a)>1$. A circuit consisting of such a location alone, with its $X$ and $Z$ components flipping different detectors, has no exact binary DEM, whereas a non-binary DEM represents it exactly.

\vspace{-2mm}
\subsection{Quantum MAP decoding and 4-cycles}
Even when both DEMs are exact (Theorem~\ref{thm:dmap}(ii)), their equivalence does not extend to decoders that estimate the fault rather than the logical class. The quantum MAP decoder maximizes over the fault variables. Thus, two DEMs for the same circuit in which fault parts have been grouped differently are in general not equivalent, and one can construct small instances in which either of the two quantum MAP decoders has the lower LER. As a consequence, BP, which approximates quantum MAP, is also expected to behave differently on the two DEMs.  
In particular, the performance of BP depends largely on the cycle structure of the Tanner graph. The following proposition formalizes the intuitive fact that splitting a location into binary parts creates 4-cycles inside the location, which do not exist in the non-binary graph.

\begin{proposition}[4-cycles inside a location]\label{prop:size}
Consider a column of rank $r$ and two detectors adjacent to it, with labels $\mathbf{h}_1$ and $\mathbf{h}_2$. Splitting the column into its binary parts creates at most $\binom{2^{r-1}}{2}$ 4-cycles through the two detectors if $\mathbf{h}_1=\mathbf{h}_2$, and at most $\binom{2^{r-2}}{2}$ otherwise, with equality if all $2^r-1$ parts have positive probability.
\end{proposition}
\begin{IEEEproof}
A part $a\in\F^{r}\setminus\{0\}$ flips detector $d$ if and only if $\langle\mathbf{h}_d,a\rangle=1$. Let $a$ be uniform on $\F^r$. Then $\langle\mathbf{h}_1,a\rangle$ and $\langle\mathbf{h}_2,a\rangle$ are uniform bits, and independent if $\mathbf{h}_1\neq\mathbf{h}_2$. Hence, $2^{r-1}$ parts flip both detectors if $\mathbf{h}_1=\mathbf{h}_2$, and $2^{r-2}$ otherwise. Any two such parts form a 4-cycle with the two detectors. Parts with zero probability are absent from the binary DEM.
\end{IEEEproof}
Note that the proposition also applies to a column obtained by regrouping, whose binary parts are the merged parts assigned to it. Its parts have distinct symptoms, hence they are distinct nonzero vectors of $\F^{r}$. For example, $\mathcal{C}'$ in Example~\ref{ex:merge} has all 15 nonzero values as parts and attains the bound. The 4-cycles between different columns are not affected.

For a 16-ary location this gives up to 28 (equal labels) or 6 (different labels) 4-cycles per pair of adjacent detectors. Table~\ref{tab:graphs} shows the effect on the circuits of Section~\ref{sec:sims}.

\begin{table}[t]
\centering
\caption{Tanner graphs of the binary DEM (Stim) and of the regrouped non-binary DEM for the circuits of Section~\ref{sec:sims} (surface and color: $d=5$, 5 rounds; BB: $[[72,12,6]]$, 6 rounds); ranks: number of binary/quaternary/8-ary/16-ary nodes.}
\label{tab:graphs}
\footnotesize
\setlength{\tabcolsep}{3pt}
\vspace{-1mm}
\begin{tabular}{@{}llrrrr@{}}
\toprule
code & DEM & VN & edges & 4-cycles & ranks\\
\midrule
surface & binary & 1\,677 & 5\,004 & 26\,046 & \\
($m=120$) & non-binary & 284 & 1\,156 & 2\,672 & 39/97/28/120\\
\midrule
color & binary & 2\,194 & 8\,617 & 186\,974 & \\
($m=90$) & non-binary & 327 & 1\,908 & 21\,089 & 72/59/0/196\\
\midrule
BB & binary & 16\,164 & 91\,116 & 2\,628\,756 & \\
($m=432$) & non-binary & 2\,016 & 17\,352 & 199\,332 & 36/432/0/1548\\
\bottomrule
\end{tabular}
\vspace{-4mm}
\end{table}

\vspace{-2mm}
\begin{figure*}[t]
\centering
\scalebox{0.95}{\includegraphics{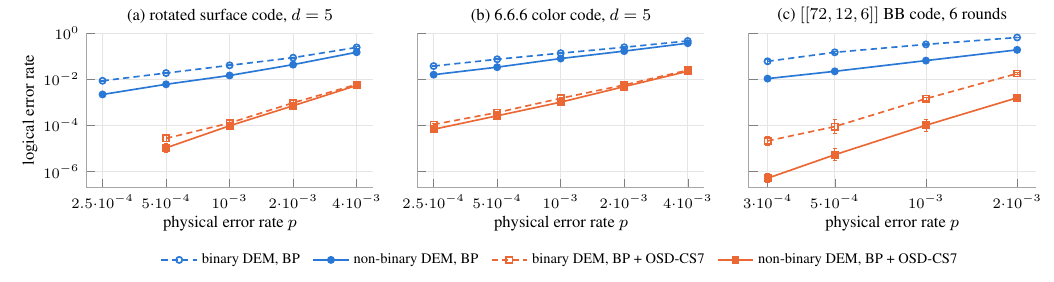}}%
\vspace{-4mm}
\caption{LER of flooded SP BP, 100 iterations, without (circles) and with OSD-CS of order 7 (squares), on the binary (dashed) and non-binary (solid) DEM under circuit-level noise of strength $p$ for $Z$-memory experiments with $d$ rounds of syndrome extraction. 95\% confidence intervals are shown.} \vspace{-4mm}
\label{fig:ler}
\end{figure*}

\vspace{-3mm}\textbf{}
\section{BP with binary messages}\label{sec:bp}
\vspace{-1mm}
The non-binary DEM can be represented as a factor graph with variable nodes of alphabet size up to 16 and \emph{binary} parity checks. Since the checks are binary, scalar messages suffice~\cite{kuo2020refined,lai2021log,ostrev2024classical,kuo2026streaming}, unlike for codes over $\mathrm{GF}(2^r)$~\cite{davey1998low}.
For every edge, we define the binary random variable $b_{d,c}=\langle\mathbf{h}_{d,c},u_c\rangle$, which can be interpreted as the contribution of column $c$ to detector $d$, so that $s_d=\sum_{c}b_{d,c}$. We write $b_{d,c}(u)$ for its value when $u_c=u$. The message on the edge between detector $d$ and column $c$ is the log-likelihood ratio (LLR) of $b_{d,c}$. Given the incoming check messages $m_{d\to c}$, variable node $c$ computes the metric
\vspace{-1mm}
\begin{equation}\label{eq:T}
T_c(u)=\log\pi_c(u)-\sum_{d}m_{d\to c}\,b_{d,c}(u),\qquad u\in\F^{r_c}, \vspace{-1mm}
\end{equation}
and sends to detector $d$ the extrinsic LLR
\vspace{-1mm}
\begin{equation}\label{eq:ext}
\lambda_{c\to d}=\maxs_{u:\,b_{d,c}(u)=0}T_c(u)-\maxs_{u:\,b_{d,c}(u)=1}T_c(u)-m_{d\to c}, %
\end{equation}
where $\maxs(a,b)=\log(e^a+e^b)=\max(a,b)+\log(1+e^{-|a-b|})$ is the Jacobian logarithm, extended to a set by applying it recursively.
The check nodes perform binary BP updates, and hence the check to variable messages are
\vspace{-2mm}
\begin{equation}\label{eq:sp} 
m_{d\to c}=(-1)^{s_d}\,2\tanh^{-1}\!\prod_{c'\in\mathcal{N}(d)\setminus c}\tanh\frac{\lambda_{c'\to d}}{2}.\vspace{-2mm}
\end{equation}
In the flooded schedule, every iteration updates all variable nodes with~\eqref{eq:ext} and then all check nodes with~\eqref{eq:sp}, starting from $m_{d\to c}=0$. The hard decision is taken per variable node (column), $\hat u_c=\argmax_{u\in\F^{r_c}}T_c(u)$, i.e., the most likely value of $u_c$. BP stops when $\sum_cH_c\hat u_c=\mathbf{s}$, and the estimate of the logical flip is $\sum_cL_c\hat u_c$. If BP does not stop within the maximum number of iterations, we declare a failure.
\textbf{Complexity.} Since BP exchanges one scalar message per edge in each direction, the non-binary DEMs of Table~\ref{tab:graphs} need $4.3$--$5.3\times$ fewer messages and check-node operations, and $5.9$--$8\times$ fewer variable-node units. In parallel hardware decoders for real-time QEC~\cite{muller2025relay}, the edges are the wires and registers between node units, and routing them is a main cost~\cite{blanksby2002690}. In exchange, variable node $c$ keeps $2^{r_c}-1$ values of $T_c(u)$ (with $T_c(0)=0$) and needs two $\maxs$ over $2^{r_c-1}$ values per edge. Counting $\deg(c)(2^{r_c}-1)$ operations per node, this is $2.3$--$2.5\times$ more than for the binary DEM.

\vspace{-1mm}
\section{Numerical results}\label{sec:sims}
\vspace{-0.5mm}
We consider $Z$-memory experiments with $d$ repeated rounds of syndrome extraction, where $d$ is the code distance, and circuit-level noise with strength $p$. More concretely, we consider a two-qubit depolarizing channel after every CNOT gate; a single-qubit depolarizing channel on every data qubit before every round (idling errors) and after every Hadamard gate, and a bit flip channel after each reset and before each measurement, all with parameter $p$ (depolarizing strength $p$ and flip probability $p$). 

We consider three QEC codes: the rotated surface code (Stim's \texttt{rotated\_memory\_z} circuit) with distance $d=5$; the distance $d=5$ 6.6.6 color code~\cite{bombin2006topological} with 19 data qubits and one $Z$- and one $X$-ancilla per plaquette; and the  $[[72,12,6]]$ bivariate bicycle (BB) code with the depth-8 syndrome cycle of~\cite{bravyi2024high} (circuit generator of~\cite{gong2024toward}). 
The binary DEM is obtained using Stim, whereas the non-binary DEM is obtained with the split-merge-regroup procedure of Section~\ref{sec:merge}\footnote{An open source implementation based on Stim is available online at \url{https://github.com/s-tinelli/nbdem}  }

As Table~\ref{tab:graphs} shows, the non-binary DEM has $5.9$--$8\times$ fewer variable nodes, $4.3$--$5.3\times$ fewer edges and $9$--$13\times$ fewer 4-cycles. For all codes, 16-ary variable nodes are the most numerous, as expected, since the split-merge-regroup procedure lets large locations absorb the parts of smaller ones.

Fig.~\ref{fig:ler} shows the LER as a function of the physical error rate $p$ for binary and non-binary DEMs under BP with a parallel (flooded) schedule with and without OSD\footnote{OSD with combination sweep of order 7~\cite{panteleev2021degenerate,roffe2020decoding}, run on the binary image of the DEM whenever BP does not converge, with columns ordered by the posterior bit LLRs and candidate cost $\sum_c-\log\pi_c(u_c)$. For the binary DEM, this is the same configuration as in \cite{roffe2020decoding}, and the OSD itself is the C++ implementation of~\cite{roffe2022ldpc}.} post-processing.
Under BP (without OSD post-processing) non-binary DEM is better on all three codes at all the $p$ values considered. For both DEMs, the LER is dominated by non-convergence, so the gain comes from an improved convergence rate.
With OSD, the non-binary DEM achieves a lower LER at every point. However, the gain is small for the surface and color codes, whereas for the BB code it grows from $11\times$ at $p=2\cdot10^{-3}$ to about $40\times$ at $p=3\cdot10^{-4}$.
The non-binary DEM also speeds up decoding. OSD is called $1.3$--$6.6\times$ less often and runs on a binary image with $2.1$--$2.3\times$ fewer columns ($r_c$ per column $c$ instead of up to $2^{r_c}-1$ parts). In our implementation, BP with OSD is $2.8$--$4.2\times$ faster on the surface and color codes and $7.5$--$9.4\times$ faster on the BB code in our CPU implementation (AMD EPYC 9274F). %
\vspace{-2mm}
\section{Conclusions}\label{sec:concl}
\vspace{-1mm}
Non-binary DEMs are equivalent to binary DEMs under degenerate decoding when the channels factorize, not under quantum MAP decoding, and have much smaller Tanner graphs, with $4.3$--$5.3\times$ fewer edges and without the 4-cycles that splitting creates inside a location. With scalar messages, they improve flooded sum-product BP, the lowest-latency schedule, by up to $6.7\times$ on surface, color and BB codes. The advantage persists with OSD post-processing, with LER reductions of up to $2.6\times$ and $1.6\times$ on the surface and color codes and of $11$--$40\times$ on the BB code, while running $2.8$--$9.4\times$ faster. Next steps are serial schedules, other post-processing methods and schedules tailored to non-binary nodes.

\FloatBarrier

\section*{Use of Artificial Intelligence}
GLM~5.3 was used for assistance in the implementation of the Stim patch that computes the non-binary
detector error models of Section~III.
The authors verified the implementation and all reported numerical results
and take full responsibility for them. 
GLM~5.3, DeepSeek~V4.1~Flash and
Claude~Opus~5.5 were used to improve the language of the manuscript. The
authors reviewed all generated text and take full responsibility for its
content.

\bibliographystyle{IEEEtran}
\bibliography{refs}

\end{document}